\PassOptionsToPackage{dvipsnames}{xcolor}
\documentclass[sigconf,nonacm]{aamas}

\usepackage{graphicx}

\usepackage{amsmath}
\usepackage{stmaryrd}
\usepackage{wasysym}

\usepackage{adjustbox}
\usepackage[inline]{enumitem}
\usepackage{url}
\usepackage{doi}
\usepackage{acronym}
\usepackage{mathtools}
\usepackage{enumitem}

\usepackage{comment}

\usepackage{soul}

\usepackage{wrapfig}

\usepackage{adjustbox}
\usepackage{caption}
\usepackage{subcaption}

\usepackage{xcolor}

\usepackage{multirow}

\usepackage{balance} % for balancing columns on the final page

\usepackage{xspace} % for \xspace

\newlength\myheight
\newlength\mydepth
\settototalheight\myheight{Xygp}
\newcommand*\inlinegraphics[1]{%
  \settototalheight\myheight{Xygp}%
  \settodepth\mydepth{Xygp}%
  \raisebox{-\mydepth}{\includegraphics[height=\myheight]{#1}}%
}

\makeatletter
\gdef\@copyrightpermission{
  \begin{minipage}{0.2\columnwidth}
   \href{https://creativecommons.org/licenses/by/4.0/}{\includegraphics[width=0.90\textwidth]{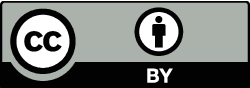}}
  \end{minipage}\hfill
  \begin{minipage}{0.8\columnwidth}
   \href{https://creativecommons.org/licenses/by/4.0/}{This work is licensed under a Creative Commons Attribution International 4.0 License.}
  \end{minipage}
  \vspace{5pt}
}
\makeatother

\setcopyright{ifaamas}
\acmConference[AAMAS '25]{Proc.\@ of the 24th International Conference
on Autonomous Agents and Multiagent Systems (AAMAS 2025)}{May 19 -- 23, 2025}
{Detroit, Michigan, USA}{Y.~Vorobeychik, S.~Das, A.~Nowé  (eds.)}
\copyrightyear{2025}
\acmYear{2025}
\acmDOI{}
\acmPrice{}
\acmISBN{}

\title[]{Decentralized Planning Using Probabilistic Hyperproperties}

\author{Francesco Pontiggia}
\affiliation{
  \institution{TU Wien}
  \city{Vienna}
  \country{Austria}}
\orcid{0000-0003-2569-6238}

\author{Filip Macák}
\affiliation{
  \institution{Brno University of Technology}
  \city{Brno}
  \country{Czech Republic}}
\orcid{0009-0004-4277-2751}

\author{Roman Andriushchenko}
\affiliation{
  \institution{Brno University of Technology}
  \city{Brno}
  \country{Czech Republic}}
\orcid{0000-0002-1286-934X}

\author{Michele Chiari}
\affiliation{
  \institution{TU Wien}
  \city{Vienna}
  \country{Austria}}
\orcid{0000-0001-7742-9233}

\author{Milan \v{C}e\v{s}ka}
\affiliation{
  \institution{Brno University of Technology}
  \city{Brno}
  \country{Czech Republic}}
\orcid{0000-0002-0300-9727}

\begin{abstract}
Multi-agent planning under stochastic dynamics is usually formalised using decentralized (partially observable) \acp{MDP} and reachability or expected reward specifications. In this paper, we propose a different approach: we use an MDP describing how a single agent operates in an environment and probabilistic hyperproperties to capture desired temporal objectives for a set of decentralized agents operating in the environment. We extend existing approaches for model checking probabilistic hyperproperties to handle temporal formulae relating paths of different agents, thus requiring the self-composition between multiple \acp{MDP}. Using several case studies, we demonstrate that our approach provides a flexible and expressive framework to broaden the specification capabilities with respect to existing planning techniques. Additionally, we establish a close connection between a subclass of probabilistic hyperproperties and planning for a particular type of \acs{Dec-MDP}s, for both of which we show undecidability. This lays the ground for the use of existing decentralized planning tools in the field of probabilistic hyperproperty verification.
\end{abstract}


\newcommand*{\llnext}[1]{\ocircle \, #1}
\newcommand*{\llglob}[1]{\square \, #1}
\newcommand*{\lleven}[1]{\Diamond \, #1}
\newcommand*{\lluntil}[2]{#1 \, \mathsf{U} \, #2}

\newcommand*{\lprobsym}{\mathbb{P}}
\newcommand*{\llprob}[1]{\mathbb{P} (#1)}

\newcommand{\mpm}{\mathcal{P}}
\newcommand{\mdpT}{(S, \Act, \mpm, L)}
\newcommand{\Act}{\mathit{Act}}
\newcommand{\act}{a}
\newcommand{\paths}{\mathit{Paths}}
\newcommand{\opaths}{\omega\mathit{Paths}}

\newcommand{\decmdp}{\mathcal{M}}
\newcommand{\dra}{\mathcal{A}}

\newcommand{\pv}{\hat{\sigma}}
\newcommand{\sv}{{\hat{s}}}
\newcommand{\quant}{\mathcal{Q}}
\newcommand{\sprod}{\mathcal{D}}
\newcommand{\spolicy}{\overline{\sigma}}

\newcommand{\reach}[3]{\mathbb{P}_{#1}({#2} \models \Diamond {#3})}

\newcommand{\distr}{\mathit{Distr}}

\newcommand{\toolname}{\textsc{PHSynt}\xspace}

\acrodef{PH}{Probabilistic Hyperproperty}
\acrodefplural{PH}{Probabilistic Hyperproperties}
\acrodef{PHL}[\textsf{PHL}]{Pro\-ba\-bi\-li\-stic Hyper Logic}
\acrodef{HYPERCTL*}[$\mathsf{HyperCTL}^{*}$]{Hyper Computation Tree Logic}
\acrodef{HYPERPCTL*}[$\mathsf{HyperPCTL}^{*}$]{Hyper Probabilistic Computation Tree Logic}
\acrodef{HYPERPCTL}[$\mathsf{HyperPCTL}$]{Hyper Probabilistic Computation Tree Logic}
\acrodef{PHLTL}[$\mathsf{PHyperLTL}$]{Probabilistic HyperLTL}
\acrodef{PHLTLDEC}[$\mathsf{PHyperLTL_{DEC}}$]{decentralised Probabilistic HyperLTL}
\acrodef{PBA}{probabilistic B{\"u}chi automaton}
\acrodefplural{PBA}[PBA]{probabilistic B{\"u}chi automata}
\acrodef{PFA}{Pro\-ba\-bi\-li\-stic Finite Automaton}
\acrodefplural{PFA}[PFA]{Pro\-ba\-bi\-li\-stic Finite Automata}
\acrodef{MC}{Markov chain}
\acrodef{MDP}{Markov decision process}
\acrodefplural{MDP}{Markov decision processes}
\acrodef{POMDP}{partially observable Markov decision process}
\acrodefplural{POMDP}{partially observable Markov decision processes}
\acrodef{Dec-MDP}{Decentralised MDP}
\acrodefplural{Dec-MDP}{Decentralised MDPs}
\acrodef{TI-Dec-MDP}{transition-independent decentralised Markov decision process}
\acrodefplural{TI-Dec-MDP}{transition-independent decentralised Markov decision processes}
\acrodef{Dec-POMDP}{decentralised partially observable MDP}
\acrodef{TM}{Turing Machine}
\acrodef{LTL}{Linear Temporal Logic}
\acrodef{LTLf}{LTL over finite traces}
\acrodef{FSA}{finite-state automaton}
\acrodefplural{FSA}[FSA]{finite-state automata}
\acrodef{DRA}{deterministic Rabin automaton}
\acrodefplural{DRA}{deterministic Rabin automata}
\acrodef{BSCC}{bottom strongly connected component}
\acrodef{MAS}{multi-agent system}
\acrodefplural{MAS}[MAS]{multi-agent systems}

\begin{document}

%%% The following commands remove the headers in your paper. For final 
%%% papers, these will be inserted during the pagination process.

\pagestyle{fancy}
\fancyhead{}

%%% The next command prints the information defined in the preamble.

\maketitle

\section{Introduction}

\paragraph{Decentralized planning under uncertainty}
\acfp{MDP} are the ubiquitous model to describe sequential decision making (or planning) of a single agent operating in an uncertain environment: the outcomes of the agent's actions are determined by a probability distribution over the successor states.
\acfp{Dec-MDP} naturally extend MDPs to the situation in which multiple agents are deployed in the same environment, each state of the environment generating a unique set of observations for the agents.
Dec-MDPs can be seen as a special case of \acfp{Dec-POMDP}~\cite{BernsteinGIZ02}, a classical model for decentralized planning under state uncertainty where the full information of the current state is not available to the agents. 
A classical synthesis task in Dec-(PO)MDPs is to compute for each agent a policy that maximises a given joint objective. Existing tool-supported approaches for Dec-(PO)MDPs typically handle only simple objectives such as time-bounded reachability or reward~\cite{Koops2023recursive} or infinite-horizon (discounted) rewards~\cite{YouTCB21}.

\paragraph{Planning of multi-agent systems using temporal logics}
In order to effectively formalise more complicated tasks and constraints, various temporal logics have been proposed in the context of \acfp{MAS}.
The most prominent are \emph{Alternating Temporal Logic} (ATL)~\cite{alur2002alternating} and \emph{Strategy Logic}~\cite{ChatterjeeHP10, MogaveroMPV14}. Extensions deal with
partial observability ~\cite{berthon2021strategy,BelardinelliLMR17,BelardinelliLMR20,catta2024obstruction}, stochastic systems ~\cite{ChenL07a,HuangSZ12,HuangL13,AminofKMMR19,BelardinelliJMM23}, and 
hyperproperties for \acp{MAS}~\cite{BeutnerF24aamas}.
\citet{BelardinelliJMM24} study model checking probabilistic variants of ATL on stochastic concurrent games with imperfect information.
Despite their expressive power, most notably the ability to express coalitions or adversarial behaviours, these logics do not specialize to the particular type of partial observability arising in \emph{decentralized} \acp{MAS}. 
%Moreover, they can express coalitions or adversarial behaviours, but, to the best of our knowledge, they support only deterministic constraints, i.e., properties that are required to hold almost surely.

\ac{LTL}~\cite{Pnueli77} has too been used to specify objectives in decentralized planning under uncertainty \cite{schuppe2021decentralized,zhu2024decomposing},
but only via manually splitting the global goal into separate, agent-local goals.

\paragraph{Probabilistic hyperproperties}
In this paper, we propose a different approach for decentralized planning under uncertainty.
The key idea is to use \acp{PH}~\cite{AbrahamBBD20,DimitrovaFT20} as the specification language. 
Hyperproperties~\cite{ClarksonS10} extend the traditional notion of trace properties by specifying requirements that put in relation multiple execution traces of a system at once.
To this extent, they consider the \emph{self-composition} of several copies of the same system, where the synchronised (step-wise paired) execution traces are analysed.
% The question of whether a trace satisfies a property depends on the other considered traces, and it can be checked on the self-composition.
For probabilistic systems, hyperproperties either relate the probability measures of different sets of traces (probabilistic computation trees) or constrain the joint probability of sets of computation trees~\cite{DOBE2022104978}.
% where, again, computation trees are considered in a synchronised fashion.
%For the latter, 
% In the verification setting~\cite{AbrahamB18,DimitrovaFT20}, each computation tree represents just a different execution of the same system.
PHs fit perfectly in a decentralized planning setting, where each different execution is interpreted as a different agent, and the hyperformula specifies a shared goal.

Our approach is inspired by pioneering studies showing that PHs have novel applications in robotics and planning \cite{DimitrovaFT20,0044NP20,AndriushchenkoBCPS23}.
In order to formalise a decentralized multi-agent planning problem, we use an MDP to describe how a single agent operates in the environment and a PH to capture relative temporal constraints on the agents.
We exemplify our approach using the following planning problem.

\begin{figure}[t]
    \centering
    \begin{subfigure}[t]{0.35\linewidth}
        \includegraphics[width=\linewidth]{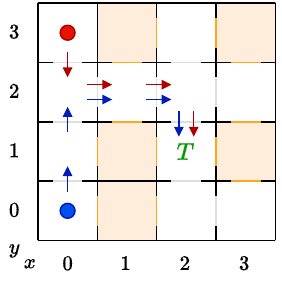}
        \caption{}
        \label{fig:running-example:memoryless}
    \end{subfigure}
    \begin{subfigure}[b]{0.28\linewidth}
        \centering
        \begin{subfigure}[t]{\linewidth}
            \centering
            \includegraphics[width=0.45\linewidth]{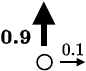}
            \caption{}
            \label{fig:running-example:slipping}
        \end{subfigure}
        \vfill
        \begin{subfigure}[t]{\linewidth}
            \centering
            \includegraphics[width=0.95\linewidth]{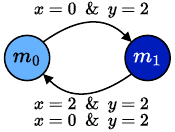}
            \caption{}
            \label{fig:running-example:controller}
        \end{subfigure}
    \end{subfigure}
    \begin{subfigure}[t]{0.35\linewidth}
        \includegraphics[width=\linewidth]{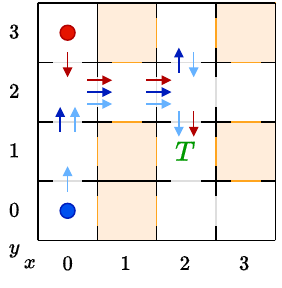}
        \caption{}
        \label{fig:running-example:memory}
    \end{subfigure}
    \vspace{-1em}
    \caption{Running example on a 4x4 grid with 2 agents and a target cell \textcolor{Green}{$T$}. The goal is to ensure that the red agent reaches \textcolor{Green}{$T$} first. (a) An illustration of the optimal memoryless policies. (b) The considered slipping factor. (c) A 2-node controller representing the optimal policy for the blue agent (the labels describes the transitions between the nodes). (d) An illustration including the optimal policy for the blue agent.}
    \vspace{-1em}
    \label{fig:running-example}
\end{figure}

\paragraph{Motivating example.}
Consider the 4x4 grid depicted in Fig.~\ref{fig:running-example:memoryless},  where obstacles are in orange.
In every white cell, an agent can move in any of the four cardinal directions and will slip right from the selected course with a probability of $0.1$ (see Fig.~\ref{fig:running-example:slipping}).  We also assume that each action can cause with some probability an irrecoverable failure. 
Consider two agents, \textcolor{blue}{$a_0$} and \textcolor{red}{$a_1$} that start in the bottom left and the upper left corners, respectively,
and seek to reach the target location $T : x=2~\&~y=1$ with maximal probability (i.e.~as soon as possible).
Additionally, we require agent~\textcolor{red}{$a_1$} to reach the target first.
The agents cannot communicate or observe the location of each other.
We can see this problem as a more complex variant of the \emph{decentralized meeting} problem~\cite{GoldmanZ04}. The specification can be encoded as a PH as follows: $\exists \, (\pv_0 \, \pv_1) \, . \, \forall \sv_{0} \in \{ s_0 \} (\pv_0) \, \forall \sv_{1}\in \{ s_1 \}(\pv_1) \, . \,
\mathbb{P}_{\max} \big((\lleven T_{\sv_0}) {\land} (\lleven T_{\sv_1}) \land \llglob (T_{\sv_0} {\implies} T_{\sv_1} )\big)$  where $s_0 : x=0~\&~y=0$ and $s_1 : x=0~\&~y=3$. Intuitively, this formula requires  two independent (i.e. decentralized) policies $\hat{\sigma}_1$ and $\hat{\sigma}_2$ for the MDP environment, one for agent $\hat{s}_0$ that starts in $s_0$, and one for agent $\hat{s}_1$ that starts in $s_1$.
The inner part requires maximising the probability that both agents reach the target, but agent~\textcolor{red}{$a_1$} arrives first.
Such decentralized requirement is not expressible in classical logics (i.e. PCTL or PLTL) because they do not allow for declaring multiple agents (via state variables $\hat{s}_i$) nor multiple policies (via policy variables $\hat{\sigma}_i$).
Fig.~\ref{fig:running-example:memoryless} shows the optimal memoryless policies.
%Our algorithm finds this pair and proves, in under one second, that no better pair (of memoryless policies) exists.
There is, however, a better memoryful strategy %that requires memory 
depicted in Fig.~\ref{fig:running-example:controller} as a 2-node controller for agent \textcolor{blue}{$a_0$}. The corresponding policy is illustrated in Fig.~\ref{fig:running-example:memory}. In this case, agent~\textcolor{blue}{$a_0$} takes a detour in cell $x=2~\&~y=2$ to give \textcolor{red}{$a_1$} more time to reach the meeting cell. This seems straightforward, however, we still want the agents to reach the target, and therefore, agent \textcolor{blue}{$a_0$} only takes the detour if they have not slipped in the cell $x=0~\&~y=2$, otherwise, they go straight to the target.

\paragraph{Contributions}
We propose a new logic, \acs{PHLTL}, as a specification language for planning in decentralized, uncertain environments.
We propose two approaches to construct decentralized policies satisfying \acs{PHLTL} specifications that both rely on an automata-based synchronised product construction.
For the full logic, we present an abstraction-refinement algorithm that constructs the optimising policy for the synchronised product MDP and attempts to factorise it into individual policies for each agent.
If factorisation is not possible, we refine the abstraction to remove such a policy from the design space and continue the search.

For a fragment of \acs{PHLTL}, the self-composition allows for translating it to a simple reachability problem for Dec-MDPs solvable by any off-the-shelf Dec-(PO)MDP solver.
As a result, we establish a close connection between a subclass of probabilistic
hyperproperties and decentralized planning.
Using a novel undecidability proof for our logic, we obtain undecidability of infinite-horizon planning in locally fully observable Dec-MDPs and thus strengthen existing undecidability results.
In the experimental evaluation, we demonstrate on a broad set of benchmarks that (i) \acs{PHLTL} can succinctly capture complex temporal constraints over policies and their executions, (ii) the proposed synthesis algorithm can solve challenging planning problems, and (iii) the algorithm, in many cases, outperforms the state-of-the-art solver on Dec-MDPs.% problems.

\subsection*{Related work}

\paragraph{Decentralized planning with complex shared objectives}

There have been several attempts at finding formalisms for expressing objectives more general than reachability in decentralized planning.

\citet{neary2021reward} use Mealy machines to encode a shared objective which is split into subtasks assigned to each agent, enabling decentralized reinforcement learning.
% In their \ac{MAS} setting, each agent keeps an \emph{abstract representation} of the other agents.
They do not use logics, but rather automata, to express requirements.

\citet{schuppe2021decentralized} use \ac{LTLf} formulae to capture safety and liveness specifications for agents modelled as \acp{MDP}.
They assume the global objective to be factored into individual agent tasks, plus a smaller task requiring their interaction.
They use stochastic games to synthesize local policies separately.
Such policies are not completely decentralized:
% if an agent cannot find an almost-surely satisfying plan autonomously,
the planner assumes that agents can share some limited information (called \emph{advisers}) on their current state, violating local observability.

\acs{LTL} specifications have been considered in the context of reinforcement learning for \acp{Dec-POMDP}:
\citet{zhu2024decomposing} propose a hierarchical approach that computes temporal equilibrium strategies
via a parity game and encode the individual reward machines~\cite{neary2021reward} used in the learning process.    

These works propose different approaches to decompose the planning problem to tackle its high complexity.
On the contrary, we perform centralized planning but tackle its complexity through abstraction refinement~\cite{CeskaJJK19,Andriushchenko022}.
Moreover, these approaches use logics to specify objectives, but aspects related to the multi-agent nature of the problem are expressed externally or in the underlying operational model because \acs{LTL} is not sufficiently expressive.

\vspace{-5pt}
\paragraph{Verification of Hyperproperties}
The deterministic hyperlogics HyperLTL and HyperCTL*~\cite{ClarksonFKMRS14}
extend \acs{LTL} and CTL* with \emph{path quantifiers} that relate multiple execution traces of the same system.
They can specify important security properties such as \emph{non-interference}~\cite{McLean94}.
\citet{BeutnerF24icaps} show that model checking of a HyperLTL fragment can be cast as a \emph{nondeterministic planning problem} for Qualitative \acp{Dec-POMDP}.
HyperSL~\cite{BeutnerF24aamas} is a variant of Strategy Logic that can relate multiple execution paths
and was used for expressing planning objectives in \acp{MAS}.
In contrast to \acs{PHLTL}, it does not support probabilistic reasoning and assumes perfect information.

\acs{PHLTL} can be seen as a probabilistic variant of HyperLTL,
that replaces path quantifiers with probability operators and quantifies over policies.
There are two other temporal logics for \acp{PH}:
\acs{HYPERPCTL}~\cite{AbrahamBBD20} and \acs{PHL}~\cite{DimitrovaFT20}.
They have been devised mainly with the purpose of verification,
while we designed \acs{PHLTL} to conveniently specify planning problems.
We compare them in Sec.~\ref{sec:related-logics}.

\textsc{HyperProb}~\cite{DobeABB21} is an SMT-based tool for model checking probabilistic hyperproperties with limited scalability~\cite{AndriushchenkoBCPS23}.
% we already say in the experiments that it does not support the self-composition
%Its scalability is limited~\cite{AndriushchenkoBCPS23}, and it does not support self-composition.
Two approaches tackle the high intractability of the problem:
\cite{DobeSBBLPW23} proposes a statistical method for bounded HyperLTL model checking on \acp{MDP} based on sampling policies and traces,
and \cite{AndriushchenkoBCPS23} introduces an abstraction refinement algorithm for the synthesis of formulae involving a comparison of probability measures for different computation trees and additional qualitative structural constraints.

\section{Background}
\label{sec:background}

A \emph{distribution} over a countable set $A$ is a~function $\mu \colon A \rightarrow [0,1]$ s.t.~$\sum_{a \in A} \mu(a) {=} 1$.
The set $\distr(A)$ contains all distributions on $A$.
Single-agent planning problems in a fully observable stochastic environment can be represented as \acp{MDP}.
\begin{definition}[MDP]
A \emph{\acf{MDP}} is a tuple $M = \mdpT$ with a finite set $S$ of states, 
a finite set $\Act$ of actions,
a transition function $\mpm \colon S \times \Act \rightarrow \distr(S)$,
and a labelling function $L \colon S \rightarrow 2^{AP}$ where $AP$ is a finite set of atomic propositions.
We denote $\mpm(s,\act,s') \coloneqq \mpm(s,\act)(s')$.
A \emph{\acf{MC}} is an \ac{MDP} with $|\Act| = 1$, denoted as $(S,\mpm, L)$ with $\mpm: S \rightarrow \distr(S)$.
\end{definition}
Note that we do not specify initial states: these will be derived from the specification.
A finite \emph{path} of an MDP $M$ is a sequence $\pi = s^0a^0s^1a^1 \dots s^n$ where ${\mpm(s^i,a^i,s^{i+1}) > 0}$ for $0 \leq i < n$.
$\paths^M$ denotes the set of all finite paths of~$M$.

A deterministic \emph{policy} (or \emph{controller}, \emph{scheduler}) is a function $\sigma \colon \paths^M \rightarrow \Act$.
Policy $\sigma$ is \emph{memoryless} if the action selection only depends on the last state of the path, i.e.~$\sigma$ maps each state $s \in S$ to an action~$\sigma(s)$.
% $\last(\pi) = \last(\pi')$ implies $\sigma(\pi) = \sigma(\pi')$ for all $\pi,\pi' \in \paths^M$.
% A memoryless policy $\sigma$ maps a state $s \in S$ to action $\sigma(s)$.
%
Policy $\sigma$ for MDP $M=\mdpT$ \emph{induces} an MC
$M^\sigma = (\paths^M, \mpm^\sigma, L^{\sigma})$
where for all paths $\pi \in \paths^M$ ending with state $s$ we set
$\mpm^\sigma(\pi,\pi a s') = \mpm(s,a,s')$ if $a = \sigma(\pi)$ and 0 otherwise,
and $L^{\sigma}(\pi) = L(s)$.

For an \ac{MDP} $M$ and a policy $\sigma$, $\reach{M^\sigma}{s}{T}$ denotes the probability of reaching from state $s \in S$ some state in a set $T \subseteq S$ of target states. 
Standard techniques~\cite{Baier2018} can efficiently compute policy $\sigma_{\max}$ that maximises this probability\footnote{We do not consider rewards here but our implementation supports them.}.
% i.e., for any policy $\sigma$ we have % $\forall \sigma \in \Sigma^M:$ 
% \mbox{$\reach{M^{\sigma_{\max}}}{s}{T}$} $\geq \reach{M^{\sigma}}{s}{T}$.
 %We formalise our approach only for the reachability probability, but our implementation supports also rewards.

To represent several agents acting in the same environment defined as an \ac{MDP},
we introduce the \ac{MDP} self-composition.
In the following, given a set $A$, we denote its $k$-ary set product as $A^k = \bigtimes_{i=1}^k A$,
and its element tuples in bold face as $\mathbf{a} = (a_1, \dots, a_k)$.

\begin{definition}[\ac{MDP} self-composition]
\label{def:self-composition}
Given a positive integer~$m$ and an \ac{MDP} $M = \mdpT$,
its \emph{$m$-self-composition} is an \ac{MDP}
$M^m = (S^m, \Act^m, \mpm^m, L^m)$ where
% $S^m$ and $\Act^m$ are $m$-ary set products, while
$\mpm^m \colon S^m \times \Act^m \rightarrow \distr(S^m)$ is s.t.\
\(
\mpm^m(\mathbf{s}, \mathbf{a}, \mathbf{s}') =
\prod_{i=1}^m \mpm(s_i, a_i, s'_i),
\)
and $L^m \colon S^m \rightarrow (2^{AP})^m$
where $L^m(\mathbf{s}) = (L(s_1), \dots, L(s_m))$.
% and $R^m : S^m \rightarrow \mathbb{R}^m$
% is s.t.\ $R^m(s_1, \dots, s_m) = (R(s_1), \dots, R(s_m))$.
\end{definition}
Intuitively, $m$-self-composition~$M^m$ describes $m$ independent instances (so-called \emph{replicas}) of $M$ that evolve simultaneously.
The notions of paths, policies, and reachability probability trivially derive from those defined for \acp{MDP}.
On the other hand, stochastic decentralized planning problems are usually modeled as \emph{decentralized POMDPs (Dec-POMDPs)}, where several agents jointly interact with a stochastic environment that they only partially observe.
% A reward function defines the system's objectives.
% In this paper, we are only interested in reachability objectives
In this paper, we assume agents can fully observe their state but not the state of other agents.
Thus, we introduce \emph{decentralized MDPs}: the definition is derived from a standard \ac{Dec-POMDP} definition (cf.~\cite{BernsteinGIZ02,BeckerZLG04}) stripped of observations and rewards.

\begin{definition}%[Dec-MDP]
A \emph{factored, observation-independent, locally fully observable \acf{Dec-MDP}} with $m$ agents is a tuple $\decmdp = \mdpT$ where
the state space $S = \bigtimes_{i=1}^m S_i$ can be factored into $m$ sets of agent-local states,
$\Act = \bigtimes_{i=1}^m \Act_i$ is the set of \emph{joint actions}, where each $\Act_i$ is a set of actions available to agent $i$, and $\mpm$ and $L$ are as in \acp{MDP}.
% $\Omega = \bigtimes_{i=1}^m \Omega_i$
% is the set of \emph{joint observations}, and each $\Omega_i$ is the finite set of observations of agent $i$;
% $s_0 \in S$ is the initial state;
% $\mpm : S \times \Act \rightarrow \distr(S)$ is the transition function,
% and $\mpm(\mathbf{s}, \mathbf{a}, \mathbf{s}')$ is the probability
% of transitioning to $\mathbf{s}'$ when action $\mathbf{a}$ is performed in state $\mathbf{s}$.
% $R : S \times \Act \times S \rightarrow \mathbb{R}^k$ is the \emph{reward} function,
% and $R(s, a, s')$ is the reward obtained when action $a$ performed in state $s$ leads to state $s'$.
% $O : S \times A \times S \rightarrow \distr(\Omega)$
% is the \emph{observation} function,
% and $O(s, a, s', o)$ is the probability of making observation $o$
% when action $a$ performed in state $s$ leads to state $s'$.
% States are \emph{jointly fully observable}, i.e.,
% if $O(s, a, s', o) > 0$ and $O(s, a, s'', o) > 0$, then $s' = s''$.
\end{definition}

% \slimparagraph{Subclasses}\textbf{(\cite{BernsteinGIZ02,GoldmanZ04,BeckerZLG04})}.
% A \ac{Dec-POMDP} is a \ac{Dec-MDP} iff it is \emph{jointly fully observable}, i.e.,
% if $O(s, a, s', o) > 0$ implies $P(s' \mid o) = 1$
% (where $P(s' \mid o)$ is the probability that a transition observing $o$ leads to state $s'$).
% Furthermore, a \ac{Dec-POMDP} is \emph{factored} if there exist $S_1, \dots, S_n$ such that $S = \bigtimes_{i=1}^n S_i$.
% Let $\decmdp$ be a factored \ac{Dec-POMDP}.
% $\decmdp$ is \emph{transition independent} if for all $i \in \mathcal{I}$
% there exists $T_i : S_i \times A_i \times S_i \rightarrow \mathbb{Q}$
% such that for all $s, s' \in S$ and $a \in A$ we have $T(s, a, s') = \prod_{i=1}^n T_i(s_i, a_i, s'_i)$.
% $\decmdp$ is \emph{observation independent} if for all $i \in \mathcal{I}$
% there exist $O_i : S_i \times A_i \times S_i \times \Omega_i \rightarrow \mathbb{Q}$
% such that for all $s, s' \in S$ and $a \in A$ we have $O(s, a, s', o) = \prod_{i=1}^n O_i(s_i, a_i, s'_i, o_i)$.
% $\decmdp$ is \emph{locally fully observable} if for all $i \in \mathcal{I}$
% there exists a function $L_i : \Omega_i \rightarrow S_i$.
% $\decmdp$ is \emph{reward independent} if there exist a monotonic function $f : \mathbb{Q}^n \rightarrow \mathbb{Q}$
% and $R_i : S_i \times A_i \times S_i$ for all $i \in \mathcal{I}$
% such that $R(s, a, s') = f(R_1(s_1, a_1, s'_1), \dots, R_n(s_n, a_n, s'_n))$.

The $m$-self-composition $M^m$ is a \ac{Dec-MDP}.
A (deterministic, me\-mo\-ry\-less) policy for agent $i$ is a mapping $\sigma_i \colon S_i \rightarrow \Act_i$, and a \emph{joint policy} $\sigma$ is a tuple $\mathbf{\sigma} = (\sigma_1, \dots \sigma_m)$ of individual policies.
For general (i.e., not locally fully observable) \acp{Dec-MDP}, the problem of finding a joint policy that maximises the probability of reaching a target state is undecidable~\cite{BernsteinGIZ02}.
We will strengthen this undecidability result to the class of locally fully observable \acp{Dec-MDP} in Sec.~\ref{sec:decModels}.

\subsection{Temporal Logic}
\label{sec:temporal-logic}

The syntax of \ac{LTL}~\cite{0020348} is given as 
\[
\varphi \coloneqq a
    \mid \varphi \land \varphi
    \mid \neg \, \varphi
    \mid \llnext{\varphi}
    \mid \lluntil{\varphi}{\varphi}
    \qquad \text{where $a \in AP$}
\]
for a finite set of atomic propositions $AP$. We consider semantics over \emph{infinite traces}, i.e., infinite sequences of subsets of $AP$.
Intuitively, $a$ holds in the current trace position if~$a$ is part of the subset of $AP$ labeling it,
$\llnext \varphi$ if $\varphi$ holds in the next position,
and $\lluntil{\varphi_1}{\varphi_2}$ holds if $\varphi_1$ holds until $\varphi_2$ does.
Propositional operators $\land$ and $\neg$ have their usual semantics.
We also use the standard abbreviations $\lleven{\varphi} = \lluntil{\top}{\varphi}$,
and $\llglob \varphi = \neg \lleven \neg \varphi$
meaning that $\varphi$ will resp.\ \emph{eventually} and \emph{always} hold.
An \ac{LTL} formula $\varphi$ can be checked on a \ac{MC} by building a deterministic Rabin automaton $\dra_\varphi$
that accepts exactly traces that satisfy $\varphi$~\cite{0020348}.

\begin{definition}[DRA]
A \acf{DRA} is a tuple $\dra = (Q, \Sigma, \delta, q_0, \mathit{Acc})$ where
$Q$ is a finite set of states, $\Sigma$~is a finite input alphabet ($\Sigma = 2^{AP}$ when checking \ac{LTL} formulae),
$\delta : Q \times \Sigma \rightarrow Q$ is the transition function,
$q_0 \in Q$ is the initial state,
and $\mathit{Acc} \subseteq 2^Q \times 2^Q$ is the \emph{Rabin acceptance condition}.
\end{definition}

A \emph{run} of $\dra$ reading the infinite word $a_0 a_1 \dots \in \Sigma^\omega$
is a sequence of states $\rho = q_0 q_1 \dots \in Q^\omega$ such that $\delta(q_i, a_i) = q_{i+1}$ for all $i \geq 0$.
Run $\rho$ is accepting if there exists a pair $(L, K) \in \mathit{Acc}$
such that states in $L$ appear only finitely often in $\rho$,
while those in $K$ appear infinitely often.
$\dra$ accepts the language of all and only words whose run is accepting.
Let $\mathbb{P}_M(s \models \varphi)$ be the probability
that the traces generated by labels of a \ac{MC} $M$
starting from a state $s$ satisfy an \ac{LTL} formula $\varphi$.
Computing $\mathbb{P}_M(s \models \varphi)$ reduces to the reachability problem
of a set of states $U_\mathit{Acc}$ in the \emph{synchronised product} between $M$ and $\dra_\varphi$.
% and then identify all of its \acp{BSCC}.
% We then call $U_\mathit{Acc}$ the set of states in \acp{BSCC} that, for some $(L, K) \in \mathit{Acc}$,
% contain no states in $L$ and at least one state in $K$.
% Thus, $\mathbb{P}_M(s \models \varphi)$ is equal to $\reach{M \times \dra_\varphi}{(s, q_0)}{U_\mathit{Acc}}$,
% which can be computed with standard techniques for \acp{MC}.
We refer to \cite{0020348} for more details.

\section{\acs{PHLTL}} 
In this section, we introduce the syntax and semantics of \ac{PHLTL}. 
The logic allows for expressing probabilistic hyperproperties on \acp{MDP} 
so that their model-checking problem can be easily interpreted as a planning problem. 
% For this, we consider \emph{existential policy quantifiers} 
% ranging over the infinite set of \ac{MDP} policies. 
% As in \acs{HYPERPCTL}, we include \emph{state quantifiers}, 
% where, for convenience, instead of quantifying over the whole set of \ac{MDP} states,
% we allow for specifying a domain of initial states~\cite{AndriushchenkoBCPS23}.
% State quantifications are crucial to construct the self-composition, 
% because they 
% (i) define the number of replicas that will form the self-composition; 
% (ii) define the initial state of each replica in the self-composition,
% and (iii) bind each replica to a specific policy, possibly making different 
% replicas to follow the same policy.

%\vspace{-0.5em}
\subsection{\acs{PHLTL} Syntax}
Let $\mathcal{V}_\mathit{pol}$ and $\mathcal{V}_\mathit{state}$ be sets of policy and state variables, respectively, and let $M = \mdpT$ be an \ac{MDP}.
Formulae in \acs{PHLTL} are constructed with the syntax
%\vspace{-2em}
\begin{align*}
   \exists \, (\pv_1 \dots \pv_n) & \, . \, \quant_1  \sv_{1} \in I_1(\pv_i) \dots \, \quant_m \sv_{m} \in I_m(\pv_j)  \, . \, \phi  \\
  & \phi \coloneqq \phi \land \phi \;
\mid \neg \, \phi \;
\mid \llprob{\varphi} \bowtie c \;
\end{align*}
where $n \leq m$ and each $\pv_i$ ($\sv_i$) is a policy (state) variable from 
$\mathcal{V}_\mathit{pol}$ ($\mathcal{V}_\mathit{state}$);
$ \quant \in \{\exists, \forall\}$, $\llprob{\cdot}$ is the \emph{probabilistic operator};
$c \in [0,1]$; each $I_i \subseteq S$ is a subset of states;
and $\varphi$ is an \ac{LTL} formula over the set
$AP_{\mathcal{V}_\mathit{state}} = \{ a_{\sv} \mid a \in AP, \sv \in \mathcal{V}_\mathit{state}\}$
of atomic propositions tagged with state variables
(cf.\ Sec.~\ref{sec:temporal-logic}).

% Intuitively, a formula is composed of a quantification prefix 
% followed by a Boolean combination of probability constraints.
A \acs{PHLTL} formula starts with a prefix that asserts the existence of $n$ policies, $\pv_1,\dots,\pv_n$, that will be the interest of the planning problem.
Then, we have $m$ agent quantifiers: $\quant_i \sv_{i} \in I_i(\pv_j)$ states that agent $i$
starts in state $\sv_i$ and is controlled by policy $\pv_j$, where a possible initial state is quantified existentially or universally from some subset $I_i \subseteq S$.
Multiple initial states are useful to express e.g.~uncertainty regarding the initial state.
Note that $m > n$ implies that some agents share the same policy.
A Boolean combination $\phi$ of \emph{probability constraints} describes the desired behaviour of agents.
A probability constraint
$\llprob{\varphi} \! \bowtie \! c$
requires the probability of the set of paths that satisfy $\varphi$
\emph{in the $m$-self-composition} of the \ac{MDP} as seen by each agent
to meet $\bowtie c$.
Atomic propositions in $\varphi$ are indexed with state variables identifying the agent for which to evaluate them.
Thus, $\varphi$ can be seen as a \acs{LTL} formula over the self-composition.

A formula is \emph{well-formed} if
all policy variables are bound by a policy quantifier---%
i.e., for every occurrence of
$\pv_j$ in e.g.~$\quant_i \sv_i {\in} I_i (\pv_j)$
there is a corresponding $\exists \pv_j$---%
and each state variable~$\sv_j$ that occurs in $\varphi$ is bound to some state quantifier $\quant \sv_j$.

\ac{PHLTL} can be straightforwardly extended with a reward operator~\cite{DobeWABB22}. Also, as common in PCTL planning, 
we consider maxi/minimization objectives $\mathbb{P}_{\mathit{max}}$ ($\mathbb{P}_{\mathit{min}}$) instead of the threshold.

\vspace{-0.5em}
\subsection{Formal Semantics} 
Let $n = |\mathcal{V}_{pol}|$ be the number of policy variables. 
A \emph{policy assignment} for $\mathcal{V}_{pol}$ is a vector 
$\Sigma = ((\pv_1, \sigma_1), \dots, (\pv_n, \sigma_n))$, 
that assigns policies $(\sigma_1 \dots \sigma_n)$ 
to all variables $(\pv_1 \dots \pv_n)$. 
We denote with $\Sigma(\pv)$ 
the policy assigned to variable $\pv$.
Probability constraints are evaluated in the context
of  \ac{MDP} $M$, a policy assignment $\Sigma$, 
a sequence of $m$ policies $\eta$,
and a state $s = (s_1, \dots, s_m)$ of the $m$-self-composition.
We use $\big( \big)$ to denote the empty sequence and $\circ$ for the concatenation operator.
$M$ satisfies formula 
$\exists \, (\pv_1 \dots \pv_n) \, . \,  \Phi$ 
iff there exists a policy assignment $\Sigma$ such that 
$M, \Sigma, \big( \big), \big( \big) \vDash \Phi$, 
where the satisfaction relation $\vDash$ is defined as follows (we omit $M, \Sigma$ for clarity):
% To simplify the notation, we omit $m$, the number of policies 
% currently in $\eta$.
\begin{align*} 
& \eta, s \vDash \exists \sv \in I(\pv) \, . \, \Phi && \text{iff } \smashoperator{\bigvee_{s' \in I}} \eta \circ \Sigma(\pv), s \circ s' \vDash  \Phi[m+1/\sv] \\ % [\sv \mapsto m+1]
\end{align*}
\begin{align*} 
& \eta, s \vDash \forall \sv \in I(\pv) \, . \, \Phi && \text{iff } \smashoperator{\bigwedge_{s' \in I}} \eta \circ \Sigma(\pv), s \circ s' \vDash  \Phi[m+1/\sv] \\ % [\sv \mapsto m+1]
& \eta, s \vDash \phi_1 \land \phi_2 && \text{iff }  \eta, s \vDash \phi_1 \text{ and }  \eta, s \vDash \phi_2\\
& \eta, s \vDash \neg \phi && \text{iff }  \eta, s \nvDash \phi \\
& \eta, s \vDash  \llprob{\varphi} \bowtie c && \text{iff }  \text{Pr}\big(\{ \pi \in \opaths(M^m_{\eta}, s) \mid {M^m_{\eta}}, \pi \vDash \varphi \}\big) \bowtie c
% &&& \text{ with $m = |\eta|$}
\end{align*}
where $m = |\eta|$, and $\Phi[m+1/\sv]$ is the formula obtained
by replacing every occurrence of a tagged proposition $p_\sv$ with the index $m+1$;
${M^m_{\eta}}$ is the MC generated by applying the $m$-tuple of policies $\eta$ to the $m$-self-composition of $M$,
and $\opaths(M^m_{\eta}, s)$ is the set of its infinite paths starting from state $s$.
Given $\pi = s^0 s^1 \dots \in \opaths(M^m_{\eta}, s)$,
we set $\pi^i = s^i$, and $\pi[i] = s^i s^{i+1} \dots$.
It is well-known that sets $ \{ \pi \in \opaths(M^m_{\eta}, s) \mid {M^m_{\eta}}, \pi \vDash \varphi\big \}$,
where $\varphi$ is an \acs{LTL} formula, are measurable~\cite{0020348}.
The $\varphi$-formulae follow the usual \ac{LTL} trace semantics, which we report below:
% The satisfaction relation for path $\pi$ is defined as follows: 
\begin{align*}
    & {M^m_{\eta}}, \pi \vDash p_j & \text{ iff } & p \in L^{m}(\pi^0)(j) \\ 
    & {M^m_{\eta}}, \pi \vDash \varphi_1 \land \varphi_2 & \text{ iff } &  {M^m_{\eta}}, \pi \vDash \varphi_1 \text{ and } {M^m_{\eta}}, \pi \vDash \varphi_2 \\ 
    & {M^m_{\eta}}, \pi \vDash \neg \varphi & \text{ iff } &  {M^m_{\eta}}, \pi \nvDash \varphi\\
    & {M^m_{\eta}}, \pi \vDash \llnext{\varphi} & \text{ iff } &  {M^m_{\eta}}, \pi[1] \vDash \varphi\\ 
    & {M^m_{\eta}}, \pi \vDash \lluntil{\varphi_1}{\varphi_2} & \text{ iff } &  \exists j \geq 0 \, . \, {M^m_{\eta}}, \pi[j] \vDash \varphi_2 \, \land \\
    &&& \forall i \in [0,j) \,  . \, {M^m_{\eta}}, \pi[i] \vDash \varphi_1
\end{align*}

\subsection{Related Logics}
\label{sec:related-logics}
Policy quantification and $\lprobsym$-operators are common to \acs{PHL}~\cite{DimitrovaFT20}, \acs{HYPERPCTL}~\cite{AbrahamBBD20}, and \acs{PHLTL}.
\acs{PHL} and \acs{HYPERPCTL} also offer alternation of policy quantifiers and comparison of $\lprobsym$-operators.
Quantifier alternation is useful for specifying adversarial problems,
but in this paper we assume a cooperative environment.
Moreover, it significantly complicates planning.
Comparison of $\lprobsym$-operators could be added to \acs{PHLTL} by integrating the abstraction refinement approach by \cite{AndriushchenkoBCPS23}.
\acs{PHLTL} and \acs{PHL} allow for full \acs{LTL} formulae in $\lprobsym$-operators,
a feature particularly suitable for planning objectives. 
\acs{HYPERPCTL} supports nested $\lprobsym$-operators.
\acs{PHLTL} and \acs{PHL} exclude them since they further complicate planning without offering significant benefits. 
%(all experiments in our paper and in related work do not require nesting).
\acs{PHLTL} and \acs{HYPERPCTL} support state quantification,
which is useful to express uncertainty in initial states of the agents.
State quantification can be expressed in \ac{PHL} with nonprobabilistic constraints:
\acs{PHLTL} is thus strictly included in \acs{PHL}.

% \acs{HYPERPCTL}~\cite{AbrahamBBD20} is a similar logic that, however,
% does not support full \acs{LTL} formulae in $\lprobsym$-operators, but allows for their nesting.
% Both \acs{PHLTL} and \acs{PHL} do not support nested $\lprobsym$-operators,
% because they would significantly complicate planning.
% %and make the model checking problem immediately undecidable~\cite{BrazdilBFK06}.
% Moreover, all the experiments in our paper and in related work do not require nesting.
% $\mathsf{HyperPCTL}^{*}$~\cite{WangNBP21} extends \acs{HYPERPCTL} with full \acs{LTL},
% but it was studied only on Markov chains.

\subsection{Undecidability}

Undecidability was proved for \acs{HYPERPCTL}~\cite{DOBE2022104978} and \ac{PHL}~\cite{DimitrovaFT20}
by encoding the emptiness problem for probabilistic B\"uchi automata, which is undecidable \cite{BaierBG08}.
The proof for \acs{HYPERPCTL} does not apply to \acs{PHLTL},
because it employs a formula containing nested probability operators. %, that \acs{PHLTL} does not support.
The formula built in the proof for \ac{PHL} contains some \acs{HYPERCTL*} parts,
but it can be rewritten as a \ac{PHLTL} formula containing, however, at least \emph{two} probability constraints.
In this section, we sketch an alternative proof
that uses a \ac{PHLTL} formula with just \emph{one} probability constraint.
We thus prove the undecidability of a smaller logic fragment,
allowing us to obtain an undecidability result for locally fully observable \acp{Dec-MDP} in Sec.~\ref{sec:decModels},
which could not be easily obtained from the other proofs.

% We prove the undecidability of \acs{PHLTL} formulae 
% by a reduction from the emptiness problem for \acp{PFA}, 
% which is known to be undecidable \cite{Paz71,condon1989complexity,Freivalds81}.

\begin{theorem}
\label{thm:phltl-undecidable}
    \acs{PHLTL} model checking is undecidable.
\end{theorem}
The proof is based on a reduction of the emptiness problem for \acp{PFA},
which is undecidable \cite{NasuH69,condon1989complexity,Freivalds81},
to model checking of a \acs{PHLTL} formula on an \ac{MDP}.
A \ac{PFA} can be seen as an \ac{MDP} whose actions are input symbols,
and it accepts finite strings (i.e., sequences of actions) for which it has
a probability $> \lambda \in \mathbb{Q}$ to reach an accepting state.
The emptiness problem consists of deciding whether such a string exists.

We build an \ac{MDP} that is the union of two disjoint \acp{MDP} $M_1$, that encodes the \ac{PFA},
and $M_2$, that is a deterministic \ac{FSA} with the same input alphabet as the \ac{PFA}.
A \acs{PHLTL} formula $\Phi$ states the existence of two policies $\sigma_1$ and $\sigma_2$,
respectively controlling $M_1$ and $M_2$, which are synchronised on the actions they take.
Moreover, $\Phi$ constrains $\sigma_1$ to
reach an accepting state of the \ac{PFA}.
One single $\lprobsym$-expression states that these constraints must be true with probability $\geq \lambda$.
If two such policies exist, since $M_2$ is a deterministic \ac{FSA},
we can extract from $\sigma_2$ a string that is accepted by the \ac{PFA}.
The full proof can be found in Appendix~ \ref{appendix:undec}.

\section{Planning by Abstraction Refinement}
\label{sec:model-checking}

Assume an MDP $M$ and a well-formed \ac{PHLTL} formula containing $n$ policy variables, $m$ state variables and a single probability expression $\mathbb{P}_{\max}(\varphi)$.
In other words, we are interested in computing an $n$-tuple of policies that maximises the probability of satisfying an LTL formula $\varphi$ containing $m$ state variables.
The definition below presents the key device used to recast the satisfaction probability of $\varphi$ into a reachability probability.

\begin{definition}
[Synchronised product]
\label{def:sync-prod}
Assume a Markov decision process $M = \mdpT$, an LTL formula $\varphi$ with $m$ state variables and a DRA $\dra_{\varphi} = (Q, (2^{AP})^m, \delta, q_0, \mathit{Acc})$ accepting traces that satisfy $\varphi$.
The \emph{synchronised product} of $M$ and $\dra_{\varphi}$ is  $\sprod(M,\varphi) = (S^m \times Q, \Act^m \times \{\alpha\}, \mathcal{P}_\sprod, L_\sprod)$, where $\alpha$ is a fresh action for transitions of $\dra_{\varphi}$
and for all $\mathbf{s}, \mathbf{s}' \in S^m$, $\mathbf{a} \in \Act^m$,
$\mathcal{P}_\sprod((\mathbf{s}, q), (\mathbf{a}, \alpha), (\mathbf{s}', q')) = \mathcal{P}^m(\mathbf{s}, \mathbf{a}, \mathbf{s}')$
where $\delta(q, L^m(\mathbf{s})) = q'$.
$L_\sprod(\mathbf{s}, q) = L^m(\mathbf{s})$ for all $\mathbf{s} \in S^m$ and $q \in Q$.
\end{definition}

We will denote $\sprod(M,\varphi)$ simply as $\sprod$ whenever $M,\varphi$ are clear from the context.
Intuitively, $\sprod$ is the product of the $m$-self-compo\-sition $M^m$ with $\dra_{\varphi}$.
Each state tuple of $\sprod$ encodes a state of each of the $m$ replicas of $M$ modelling the $m$ agents, and a state of $\dra_\varphi$.
The actions of $\sprod$ synchronise $m$ actions of $M$ with the (deterministic) update of $\dra_{\varphi}$ on $m$-tuples of labels of $M$.
The theorem below asserts that computing the satisfaction probability of $\varphi$ on $M$ is equivalent to computing a reachability probability on $\sprod$. The proof follows directly from Def.~\ref{def:self-composition}, Def.~\ref{def:sync-prod} and~\cite{0020348}.

\begin{theorem}
Assume $\mathbf{s} \in S^m$ and a policy $\sigma \in \Sigma^{M^m}$. Then
\[
\mathbb{P}_{(M^m)^\sigma}(\mathbf{s} \vDash \varphi)
=  \reach{\sprod^{\sigma'}}{(\mathbf{s}, \delta(q_0, L^m(\mathbf{s})))}{U_{\dra_{\varphi}}}
\]
where $\sigma'(\pi) = (\sigma(\pi), \alpha)$ for any path $\pi$,
and $U_{\dra_{\varphi}}$ is the success set 
of $\dra_{\varphi}$'s acceptance condition.  
\end{theorem}

\subsection{Synchronised Product as an Abstraction}
Assume the \acs{PHLTL} specification from above where we seek an $n$-tuple of policies that maximises the probability of satisfying an LTL formula $\varphi$ containing $m$ state variables.
With standard methods for MDP model checking, we can compute policy $\spolicy_{\max}$ maximising $\llprob{\lleven U_{\dra_{\varphi}}}$ on $\sprod$.
$\spolicy_{\max}$ maps state tuples $(s_1, \dots, s_m, q)$ of $\sprod$ to action tuples $(a_1, \dots, a_m, \alpha)$.
Under two conditions, $\spolicy_{\max}$ can be \emph{factorised} into an $n$-tuple $(\sigma_1, \dots, \sigma_n)$ of policies for individual agents.
Intuitively, a policy $\spolicy$ for $\sprod$ is a centralised policy that makes decisions while considering the states of all agents and thus may violate the local observability assumption. 
Additionally, $\spolicy$ ignores the fact that $m$ might be larger than $n$, i.e. the case where multiple agents (replicas) must adhere to the same policy.
The following definition formalises these concepts.

\begin{definition}[Policy consistency]
\label{def:policy-inconsistency}
Let $\spolicy$ be a policy for $\sprod$.
We say that $\spolicy$ \emph{violates local observability} if for some agent $i$ there exist two states $\mathbf{s}$ and $\mathbf{s'}$ s.t.~$s_{i} = s'_{i}$ and $\spolicy(\mathbf{s})(i) \neq \spolicy(\mathbf{s}')(i)$.
We say that~$\spolicy$ \emph{violates policy bindings} if for two agents $i$ and $j$, $i \neq j$, bound to the same policy variable, there exist two (not necessarily different) states $\mathbf{s}$ and $\mathbf{s'}$ s.t.~$s_{i} = s_{j}'$  and $\spolicy(\mathbf{s})(i) \neq \spolicy(\mathbf{s}')(j)$.
We say that $\spolicy$ is \emph{inconsistent} if it violates local observability or policy bindings.
\end{definition}

We remark that every $n$-tuple of policies for $M$ can be mapped to a policy for $\sprod$, but only a consistent policy $\spolicy$ can be factorised into an $n$-tuple of policies $(\sigma_1, \dots, \sigma_n)$ as follows: $\sigma_i(s) = \spolicy(\mathbf{s})(j)$ where $j$ is the index of an agent bound to $i$-th policy $\pv_i$ and $\mathbf{s}_j = s$.
This makes $\sprod$ a proper abstraction for $M$ and $\varphi$.

\subsection{Abstraction Refinement}
\label{sec:AR}
Assume we require\footnote{$\lambda$ can be the value of the current optimum if we seek $\mathbb{P}_{\max}(\varphi)$.} $\llprob{\varphi} > \lambda$.
Since $\sprod$ is an abstraction, i.e.~$\Sigma^{\sprod}$ encompasses all policy tuples for $M$ and $\varphi$, the value $V(\spolicy_{\max})$ of the maximizing policy for $\sprod$ is an over-approximation of the maximal value achievable by a policy $n$-tuple.
If $V(\spolicy_{\max}) > \lambda$ and $\spolicy_{\max}$ is consistent, then it can be factorized into an $n$-tuple of policies that satisfy the property.
If $V(\spolicy_{\max}) \leq \lambda$, then no $n$-tuple of policies can satisfy the property.
If $V(\spolicy_{\max} > \lambda)$ and $\spolicy_{\max}$ is inconsistent, we optimistically obtain a policy by randomly resolving the inconsistencies in $\spolicy_{\max}$, and check if the resulting policy $n$-tuple satisfies the threshold. If we fail, no conclusion can be deduced. Inspired by the abstraction refinement approach of~\cite{CeskaJJK19,Andriushchenko022}, we \emph{split} $\sprod$ into sub-models, refining the abstraction and ensuring that the inconsistencies obtained for $\sprod$ are excluded from the refined model.

To split $\sprod$, we inspect the inconsistent policy $\spolicy_{\max}$.
From Def.~\ref{def:policy-inconsistency} it follows that there must exist two states $\mathbf{s},\mathbf{s}'$ and two (not necessarily different) agents $i,j$, bound to the same policy variable, for which $\mathbf{s}_i = \mathbf{s}_j'$ and such that action selections $a \coloneqq \spolicy_{\max}(\mathbf{s})(i)$ and $b \coloneqq \spolicy_{\max}(\mathbf{s}')(j)$, $a \neq b$, violate local observability or policy binding.
Let $\mathcal{H}$ be the set of all agents bound to the same policy variable as $i$ and $j$.
Splitting generates 3 submodels, where, for an arbitrary state $\mathbf{s}$ of $\sprod$ and agent $k \in \mathcal{H}$:
$\sprod_1$ with $\Act(\mathbf{s})(k) = a$;
$\sprod_2$ with $\Act(\mathbf{s})(k) = b$;
and $\sprod_3$ with $\Act(\mathbf{s})(k) = \Act(\mathbf{s}) {\setminus} \{ a, b \}$.

Upon splitting and obtaining a refined abstraction $\sprod_i$, its analysis proceeds as that of $\sprod$.
For optimality objectives, this yields a recursive anytime algorithm tracking the currently best result.

Correctness of the algorithm follows from the construction of~$\sprod$ and, in particular, the synchronised product with DRA $\dra_{\varphi}$.

\begin{theorem}
Abstraction refinement terminates and either returns a consistent tuple of memoryless policies $\sigma_1 \dots \sigma_n$ witnessing the specification or proves that no such tuple exists.
\end{theorem}

\paragraph{Complexity.} 
$|\sprod|$ is doubly exponential in the length of $\varphi$ and singly exponential in $m$.
%This complexity is optimal wrt.\ the formula~\cite{CourcoubetisY95}.
Moreover, the number of (memoryless) policies 
%$|\Sigma|$ = $(2^{|M| \cdot n})$
%, i.e., 
is exponential in both $n$ and the number of states in $M$ 
and thus naive enumeration of policies is intractable. Indeed, \ac{Dec-MDP} problems are NEXPTIME-hard ~\cite{BernsteinGIZ02}, even for a finite horizon. 

\paragraph{Beyond memoryless policies}
The presented algorithm analysing the synchronised product $\sprod$ can only find tuples of memoryless policies witnessing the specification.
As we demonstrated in the motivating example, using memory might allow agents to satisfy specifications where simple behaviour described by memoryless policies is not enough, or to achieve better values when searching for the optimal behaviour.
To enable the search for non-memoryless policies, we can unfold memory into the MDP $M$ itself, allowing each agent to perform original actions while simultaneously updating their memory, if necessary.

\paragraph{Multiple probability constraints.} 
An extension to handle multiple probability constraints derives straightforwardly from ~\cite{AndriushchenkoBCPS23}. We analyse each probability constraint in the abstraction $\sprod$ as above, and, by computing also a minimizing policy $\spolicy_{\min}$, we mark it as `SAT' (all policies satisfy the constraint), `UNSAT' (no policy), or `ambiguous' (some policy). These results are then combined according to the boolean operators. If this combination does not generate a satisfying policy we split $\sprod$ w.r.t. an ambiguous result, and recursively analyse submodels. If it generates an `UNSAT' result, we discard $\sprod$.
%and obtain refined abstractions $\sprod_i$ and $\sprod_j$.
%If this combination is `SAT', we have found a solution and the formula is satisfiable. If in some refined abstraction $\sprod_i$, the combination is `UNSAT', we can discard $\sprod_i$.
% We then combine the results according to their combining operator, to obtain a satisfying policy for the whole combination. 
% If unsuccessful, we split w.r.t. an ambiguous result.

\section{From \acs{PHLTL} to Dec-MDP planning}
\label{sec:decModels}

We introduce \acs{PHLTLDEC}, a fragment of \acs{PHLTL} that can be reduced to \ac{Dec-MDP} planning.
It consists of all formulae in which 
i) each policy is associated with exactly one agent,
ii) there is a single probability operator, and
iii) all $I_1 \dots I_m$ are singleton sets. 
\acs{PHLTLDEC} formulae have the following syntax:
\begin{center}
\(
    %\Phi_{\mathcal{DEC}} \coloneqq
    \exists (\pv_1 \dots \pv_m) . \forall \sv_{1} \in \{ s_1 \} (\pv_1) \dots \forall \sv_{m}\in \{ s_m \}(\pv_m) . \llprob{\varphi} \bowtie \lambda
\)
\end{center}
Given a \acs{PHLTLDEC} formula $\Phi$ over an MDP $M$,
it is easy to see that the synchronised product $\sprod$ described in Def.~\ref{def:sync-prod}
is a \ac{Dec-MDP} with $m+1$ agents---one for each replica of $M$ in the self-product,
plus one for the \ac{DRA}.
Given $\mathbf{s} = (s_1, \dots, s_m)$,
the tuple of initial states associated to each policy variable in $\Phi$,
$\llprob{\varphi}$ is equivalent to the probability of reaching
the acceptance set of the \ac{DRA} from $s$ in the \ac{Dec-MDP}. 
Despite the apparent simplicity of this connection, we are, to the best of our knowledge, the first to relate a class of probabilistic hyperproperties to Dec-MDP planning.

This fact has two main consequences.
First, we can exploit existing planning algorithms for \acp{Dec-MDP}
to model check the fragment.
In Sec.~\ref{sec:eval}, we compare our abstraction refinement algorithm from Sec.~\ref{sec:model-checking}
to a state-of-the-art \ac{Dec-MDP} planning tool.

Second, we derive a new undecidability result for a class of \acp{Dec-MDP}.
Since a \acs{POMDP} is a special case of \ac{Dec-MDP} \cite{BernsteinGIZ02}, 
the planning problem for general \acp{Dec-MDP} is known to be undecidable.
The construction of $\sprod$ turns the verification of a \acs{PHLTLDEC} formula on $M$ into a planning problem for a \emph{locally fully observable} \ac{Dec-MDP}. 
%A \ac{PFA} is also a locally fully observable \ac{Dec-MDP},
So from Theorem~\ref{thm:phltl-undecidable} follows that planning is undecidable for \acp{Dec-MDP}
even if they are locally fully observable:
\begin{corollary}
    The planning problem for locally fully observable \acp{Dec-MDP} under the infinite-horizon total reward criterion is undecidable. 
\end{corollary}

\section{Experimental Evaluation}
\label{sec:eval}
We implemented the proposed abstraction refinement approach in the tool \toolname. The tool is built on top of PAYNT~\cite{AndriushchenkoC21}, a tool for verification approaches for families of MCs, using
STORM~\cite{STORM} to solve MDP model-checking queries 
and SPOT~\cite{Duret-LutzRCRAS22} to generate Rabin automata for LTL formulae. 
The implementation and all the considered benchmarks are publicly available\footnote{\url{https://github.com/francescopont/hyper-synthesis}}.
The evaluation focuses on the following questions:

\begin{itemize}[leftmargin=15pt]
    \item Can \ac{PHLTL} capture relevant  planning specifications? 
    \item Can \toolname adequately solve such problems?
    \item Can \toolname find even better solutions by searching for non-memoryless policies?
    \item How does the performance of \toolname fare against state-of-the-art Dec-PMDP solver Inf-JESP~\cite{YouTCB21} for \acs{PHLTLDEC}?
\end{itemize}

We could not compare with the model-checking tool \textsc{HyperProb}~\cite{DobeABB21} because it does not support the self-composition.

\subsection{Experimental setting} 
\label{sec:setting}

\paragraph{Benchmarks} 
We consider decentralized planning problems over a grid-world environment~\cite{hauskrecht1997incremental} with various obstacles.
Agents move in the four cardinal directions (via four actions), and each action has a small probability of transitioning into an unintended neighbouring cell. 
We consider four variants of different dimensions and varying complexity of the specification. The particular specifications are described case by case later. In all models, we equip each action with a non-zero probability that the agent transitions to a \emph{trap state}. This allows us to explicitly model \emph{discounting} and thus conduct a fair comparison with Inf-JESP that currently does not support undiscounted specifications.

\paragraph{Main tables} For every model, we report experimental results in Tables~\ref{tab:results-dec} and~\ref{tab:results-other}. Here $|\decmdp|$ denotes the size of the underlying MDP, i.e. the state space of one agent. We report \emph{bounds} on the values that can be achieved for the respective planning problem. A baseline is the value of a randomized policy (in each state, all agents uniformly pick one of the available actions), and an upper bound is given by solving the product MDP $\sprod$, i.e. assuming a centralized planning problem where each agent has perfect information about other agents. The table further shows the size of the product $|\sprod|$ for memoryless policies ($mem=0$) and for policies using 1-bit memory ($mem=1$).
For \toolname (and Inf-JESP in Table~\ref{tab:results-dec}), we report the runtimes (in seconds) and the best values achieved.
$\dagger$ indicates that \toolname was not able to explore the whole search space: we report the best value found within a time limit of 1 hour, and the time when the best value was found.
When running the experiment with $mem=1$, a value obtained for $mem=0$ is given to the synthesiser as a reference point. The goal here is to see whether adding memory can help the synthesiser find  a better value: $\dagger$ without the number in parentheses indicates that no better policy has been found.

\paragraph{Using Inf-JESP} Inf-JESP leverages point-based belief exploration methods, in particular, it repeatedly calls SARSOP~\cite{KurniawatiHL08} on sub-problems given as (single agent) POMDPs. Inf-JESP builds on randomised initialization and individual runs of the tool can vary massively. Inspired by the experimental evaluation of Inf-JESP presented in~\cite{YouTCB21}, we use 10 restarts and run the tool 30 times with an overall time limit of 2 hours. It means that we get at most 300 individual runs per experiment. We report the overall runtime of the experiment\footnote{The runtime can exceed 2 hours since we do not terminate runs that already started.} and the best value~achieved.

\subsection{Results}

\begin{table*}[t]
\captionsetup{font=small}
\caption{Results for decentralised planning problems that can be encoded as Dec-MDPs. All models maximise, meet$^R\!$ minimises, hence its bound is actually a lower bound. Values in bold indicate the best results. Individual columns are described in Sec.~\ref{sec:setting}.}
\vspace{-1em}
\label{tab:results-dec}

\setlength{\tabcolsep}{1pt}
\renewcommand{\arraystretch}{0.8}
\scalebox{0.9}{
\begin{tabular}{
l r@{\hskip 12pt}cc@{\hskip 12pt}cr@{\hskip 12pt}rr@{\hskip 12pt} rr@{\hskip 24pt}
l r@{\hskip 12pt}cc@{\hskip 12pt}cr@{\hskip 12pt}rr@{\hskip 12pt} rr
}
\toprule

\multirow{2}{*}{\textbf{model}} &
\multirow{2}{*}{$|\decmdp|$} &
\multicolumn{2}{c@{\hskip 12pt}}{bounds} &
\multicolumn{2}{c@{\hskip 12pt}}{product} &
\multicolumn{2}{c@{\hskip 12pt}}{\toolname} &
\multicolumn{2}{c@{\hskip 24pt}}{Inf-JESP} &
\multirow{2}{*}{\textbf{model}} &
\multirow{2}{*}{$|\decmdp|$} &
\multicolumn{2}{c@{\hskip 12pt}}{bounds} &
\multicolumn{2}{c@{\hskip 12pt}}{product} &
\multicolumn{2}{c@{\hskip 12pt}}{\toolname} &
\multicolumn{2}{c}{Inf-JESP}\\

\cmidrule(lr{1.75em}){3-4}
\cmidrule(lr{1.75em}){5-6}
\cmidrule(lr{1.75em}){7-8}
\cmidrule(lr{3.25em}){9-10}
\cmidrule(lr{1.75em}){13-14}
\cmidrule(lr{1.75em}){15-16}
\cmidrule(lr{1.75em}){17-18}
\cmidrule(lr){19-20}
& & rand. & upp. & mem & $|\mathcal{D}|$ & time & value & time & value &
& & rand. & upp. & mem & $|\mathcal{D}|$ & time & value &  time & value  \\

\midrule
\multirow{2}{*}{meet-4x4} & \multirow{2}{*}{22} & \multirow{2}{*}{0.1} & \multirow{2}{*}{0.66} & 0 & 122 & <1 & 0.63 & \multirow{2}{*}{15} & \multirow{2}{*}{0.63} & 
\multirow{2}{*}{meet-12x7} & \multirow{2}{*}{78} & \multirow{2}{*}{0.04} & \multirow{2}{*}{0.66} & 0 & 2k & 347 & 0.63 & \multirow{2}{*}{1839} & \multirow{2}{*}{\textbf{0.64}}   \\ 
& & & & 1 & 485 & $\dagger$ & 0.63 &
& & & & & & 1 & 6k & $\dagger$ & 0.63  \\ 

\midrule
\multirow{2}{*}{meet-15x9} & \multirow{2}{*}{160} & \multirow{2}{*}{0} & \multirow{2}{*}{0.47} & 0 & 6k & $\dagger$ (41) & 0.32 & \multirow{2}{*}{351} & \multirow{2}{*}{0} & 
\multirow{2}{*}{meet-19x11} & \multirow{2}{*}{216} & \multirow{2}{*}{0.01} & \multirow{2}{*}{0.57} & 0 & 12k & $\dagger$ (9) & 0.29  & \multirow{2}{*}{7298} & \multirow{2}{*}{0.17}   \\ 
& & & & 1 &  26k & $\dagger$(370) & \textbf{0.33} &
& & & & & & 1 & 47k & $\dagger$ (47) & \textbf{0.56} \\

\midrule
\multirow{2}{*}{meet$^R\!$-4x4} & \multirow{2}{*}{22} & \multirow{2}{*}{322} &\multirow{2}{*}{106} & 0 & 122 & <1 & 117 & \multirow{2}{*}{1262} & \multirow{2}{*}{118} &
\multirow{2}{*}{meet$^R\!$-12x7} & \multirow{2}{*}{78} & \multirow{2}{*}{347} & \multirow{2}{*}{104} & 0 & 2k & 9 & \textbf{109} & \multirow{2}{*}{4572} & \multirow{2}{*}{128} \\
& & & & 1 & 485 & $\dagger$ (355) & \textbf{114} &
& & & & & & 1 & 6k & $\dagger$ & 109  \\

\midrule
\multirow{2}{*}{meet$^R\!$-15x9} & \multirow{2}{*}{160} & \multirow{2}{*}{43} & \multirow{2}{*}{21} & 0 & 6k & $\dagger$ (498) & \textbf{22} & \multirow{2}{*}{509} & \multirow{2}{*}{25} &
\multirow{2}{*}{meet$^R\!$-19x11} & \multirow{2}{*}{216} & \multirow{2}{*}{655} & \multirow{2}{*}{240} & 0 & 12k & 421 & \textbf{245}  & \multirow{2}{*}{9730} & \multirow{2}{*}{606} \\
& & & & 1 & 26k & $\dagger$ & 22 &
& & & & & & 1 & 47k & $\dagger$ & 245  \\

\midrule
\multirow{2}{*}{race-2-4x4} & \multirow{2}{*}{18} & \multirow{2}{*}{0.2} & \multirow{2}{*}{0.8} & 0 & 76 & <1 & 0.7 & \multirow{2}{*}{9} & \multirow{2}{*}{0.72} &
\multirow{2}{*}{race-2-12x7} & \multirow{2}{*}{64} & \multirow{2}{*}{0.07} & \multirow{2}{*}{0.7} & 0 & 236 & 1 & \textbf{0.65}  & \multirow{2}{*}{1457} & \multirow{2}{*}{0.62}  \\
& & & & 1 & 295 & 2777 & \textbf{0.76} &
& & & & & & 1 & 935 & $\dagger$ & 0.65  \\

\midrule
\multirow{2}{*}{race-2-15x9} & \multirow{2}{*}{160} & \multirow{2}{*}{0.02} & \multirow{2}{*}{0.58} & 0 & 6k & 26 & \textbf{0.58} & \multirow{2}{*}{195} & \multirow{2}{*}{0.51} &
\multirow{2}{*}{race-2-19x11} & \multirow{2}{*}{168} & \multirow{2}{*}{0.01} & \multirow{2}{*}{0.61} & 0 & 2k & 19 & \textbf{0.57}  & \multirow{2}{*}{2968} & \multirow{2}{*}{0.53}  \\
& & & & 1 & 26k & $\dagger$ & 0.58 &
& & & & & & 1 & 7k & $\dagger$ & 0.57  \\

\midrule
\multirow{2}{*}{race-3-4x4} & \multirow{2}{*}{18} & \multirow{2}{*}{0.17} & \multirow{2}{*}{0.76} & 0 & 228 & 0 & 0.65 & \multirow{2}{*}{-} & \multirow{2}{*}{-} &
\multirow{2}{*}{race-3-12x7} & \multirow{2}{*}{64} & \multirow{2}{*}{0.02} & \multirow{2}{*}{0.61} & 0 & 2k & $\dagger$ & 0.42  & \multirow{2}{*}{-} & \multirow{2}{*}{-} \\
& & & & 1 & 2k & $\dagger$ (346) & 0.71 &
& & & & & & 1 & $\dagger$ & $\dagger$  & $\dagger$ \\

\midrule
\multirow{2}{*}{race-3-15x9} & \multirow{2}{*}{160} & \multirow{2}{*}{0} & \multirow{2}{*}{0.39} & 0 & 513k & $\dagger$ (528) & 0.27 & \multirow{2}{*}{-} & \multirow{2}{*}{-} &
\multirow{2}{*}{opac-4x4} & \multirow{2}{*}{42} & \multirow{2}{*}{0} & \multirow{2}{*}{0.37} & 0 & 529 & 5 & 0.09 & \multirow{2}{*}{80} & \multirow{2}{*}{\textbf{0.15}}
\\
& & & & 1 & $\dagger$ & $\dagger$ & $\dagger$ &
& & & & & & 1 & 2k & $\dagger$ (1000) & 0.14
\\

\midrule
\multirow{2}{*}{opac-12x7} & \multirow{2}{*}{130} & \multirow{2}{*}{0} & \multirow{2}{*}{0.45} & 0 & 6k & 25 & 0.11 & \multirow{2}{*}{3730} & \multirow{2}{*}{\textbf{0.19}} &
\multirow{2}{*}{opac-15x9} & \multirow{2}{*}{540} & \multirow{2}{*}{0} & \multirow{2}{*}{0.29} & 0 & 94k & $\dagger$ (3) & \textbf{0.01} & \multirow{2}{*}{2234} & \multirow{2}{*}{0}
\\
& & & & 1 & 23k & $\dagger$ (349) & 0.15 &
& & & & & & 1 & 376k & $\dagger$ & 0.01 & \\

\bottomrule
\end{tabular}
}
\end{table*}

\setlength{\textfloatsep}{15pt}
\begin{table}[t]
\captionsetup{font=small}
\caption{
The results for the decentralised planning problems that can not be encoded as Dec-MDPs. All models maximise. %The individual columns are described in Sec.~\ref{sec:setting}.
}
\vspace{-1em}
\label{tab:results-other}

\setlength{\tabcolsep}{4pt}
\renewcommand{\arraystretch}{0.8}
\scalebox{0.95}{
\begin{tabular}{l rcc@{\hskip 12pt}rr@{\hskip 12pt}rr}
\toprule

\multirow{2}{*}{\textbf{model}} &
\multirow{2}{*}{$|\decmdp|$} &
\multicolumn{2}{c@{\hskip 12pt}}{bounds} &
\multicolumn{2}{c@{\hskip 12pt}}{product} &
\multicolumn{2}{c}{\toolname} \\

\cmidrule(lr{1.75em}){3-4}
\cmidrule(lr{1.75em}){5-6}
\cmidrule(lr){7-8}
& & rand. & upp. & mem & $|\mathcal{D}|$ & time & value  \\

\midrule 
\multirow{2}{*}{ISO-4x4} & \multirow{2}{*}{22} & \multirow{2}{*}{0.01} & \multirow{2}{*}{0.47} & 0 & 1k & <1 & 0.08 \\
& & & & 1 & 4k & 1677  & 0.15 \\

\midrule 
\multirow{2}{*}{ISO-12x7} & \multirow{2}{*}{78} & \multirow{2}{*}{0.01} & \multirow{2}{*}{0.46} & 0 & 13k & 6 & 0.2 \\
& & & & 1 & 51k & $\dagger$ & 0.2 \\

\midrule 
\multirow{2}{*}{ISO-15x9} & \multirow{2}{*}{160} & \multirow{2}{*}{0.02} & \multirow{2}{*}{0.29} & 0 & 74k & $\dagger$(2244)  & 0.04 \\
& & & & 1 & 297k & $\dagger$ & 0.04 \\

\midrule
\multirow{2}{*}{ISO-19x11} & \multirow{2}{*}{216} & \multirow{2}{*}{0.01} & \multirow{2}{*}{0.37} & 0 & 98k & 1093 & 0.09 \\
& & & & 1 & 390k  & $\dagger$  & 0.09 \\

\midrule 
\multirow{2}{*}{robust-4x4} & \multirow{2}{*}{35} & \multirow{2}{*}{0.23} & \multirow{2}{*}{0.7} & 0 & 281 & <1 & 0.64 \\
& & & & 1 & 1k & $\dagger$ & 0.64 \\

\midrule             
\multirow{2}{*}{robust-12x7} & \multirow{2}{*}{128} & \multirow{2}{*}{0.12} & \multirow{2}{*}{0.63} & 0 & 375 & <1 & 0.59  \\
& & & & 1 & 1k & $\dagger$ & 0.59 \\

\midrule
\multirow{2}{*}{robust-15x9} & \multirow{2}{*}{640} & \multirow{2}{*}{0.04} & \multirow{2}{*}{0.59} & 0 & 23k & $\dagger$ (13) & 0.53  \\
& & & & 1 & 91k & $\dagger$ & 0.53 \\

\midrule
\multirow{2}{*}{robust-19x11} & \multirow{2}{*}{203} & \multirow{2}{*}{0.14} & \multirow{2}{*}{0.73} & 0 & 9k & 15 & 0.64 \\
& & & & 1 & 36k & $\dagger$ & 0.64 \\

\midrule
noninter-4x4 & 22 & 0.25 & 0.71 & 0 & 51k & 80 &  0.69  \\

\bottomrule
\end{tabular}
}
\vspace{-0.5em}
\end{table}

We present 6 planning problems, demonstrate their specification via probabilistic hyperproperties, and discuss the solutions found by our approach.
The first three problems are within the fragment $\acs{PHLTLDEC}$, so we will also evaluate these using Inf-JESP. All the experiments are run on a device equipped with AMD Ryzen 5 6600U @2.9GHz and 64GB RAM.

\paragraph{Meeting} 
We consider a classical decentralized planning problem~\cite{GoldmanZ04} where two agents, each starting from different initial locations $s_1$ and $s_2$, seek to meet at a given target location $T$.
%Every step is associated with a small probability of crashing, causing an irreparable error of an agent.
Note that the two agents do not have the option to sit still upon reaching the target, and therefore the probability of both agents residing at the same location is not the product of individual reachability probabilities.
%The problem is close to the class of \emph{Goal-Oriented Dec-MDPs}~\cite{GoldmanZ04}, a subclass of Dec-MDPs admitting a polynomial-time algorithm for finite-horizons.
% In GO-Dec-MDPs, agents earn a positive reward in some goal states, while everywhere else they incur a cost, given by summing the costs of the two agents' local states.
In the following \acs{PHLTL} formula, we require to maximise the probability of meeting, thus to meet as soon as possible (recall the transitions to the trap state)
\begin{center}
    $\exists \, (\pv_1 \, \pv_2) \, . \, \forall \sv_{1} \in \{ s_1 \} (\pv_1) \, \forall \sv_{2}\in \{ s_2 \}(\pv_2) \, . \, \mathbb{P}_{\max} (\lleven \, (T_{\sv_1} \land T_{\sv_2}))$
\end{center}
We observe that, for smaller grids, memoryless controllers obtained with \toolname often suffice to get a close-to-maximal probability: often simply heading to the meeting location is good enough.
Inf-JESP achieves similar results, even marginally outperforming \toolname at meet-12x7.
For larger grids, however, the value achieved by the memoryless policy gets further from the upper bound: memory is needed to navigate different paths that facilitate the meeting.
In those cases, \toolname manages to improve the achieved probability when policies are equipped with 1 bit of memory.
On the other hand, Inf-JESP struggles to find solutions that outperform memoryless policies.
Due to the exponential blow-up of the search space (and the increased size of the model), \toolname does not fully explore the design space of policies within the timeout, although the improved solution is usually obtained very quickly, e.g.~in 47 seconds for meet-19x11.

We also consider a variant meet$^R$ where the goal is for the two agents to meet at a designated area while minimising costs associated with the movement of one of the agents. This variant was designed to include in our benchmark suite also problems with non-sparse rewards.
We observe that even memoryless policies are sufficient to achieve close-to-optimal behaviour.
\toolname again often produces close-to-optimal solutions. This does not mean that these problems are trivial, cf.~meet$^R\!$-19x11, where Inf-JESP cannot find adequate policies even after 2 hours.

\paragraph{Race}
We consider the specification from the 
motivating example (see the introduction)
%here the example from Fig.\ref{fig:running-example}. 
%In \acs{PHLTL}, 
%\begin{gather*}
%    \exists \, (\pv_1 \, \pv_2) \, . \, \forall \sv_{1} \in \{ s_1 \} (\pv_1) \, \forall \sv_{2}\in \{ s_2 \}(\pv_2) \, . \, \\
%    \llmaxProb{(\lleven \, \, T_{\sv_1}) \land (\lleven T_{\sv_2}) \land \llglob (T_{\sv_1} \implies T_{\sv_2} )}
%\end{gather*}
that can be seen as a probabilistic variant of the \emph{shortest path} hyperproperty~\cite{0044NP20}.
%, which requires to synthesize a path (in a non-probabilistic environment) that reaches the target before any other path.
The race-3 variant reported in Table~\ref{tab:results-dec} considers three agents with a specific desired order of arrival.
The obtained policies indicate again that often memoryless policies are good enough, depending on the initial locations of the agents that may already favour the intended winner of the race.
%Nonetheless, for race-2-4x4, \toolname manages to find the (better) solution of Fig.~\ref{fig:running-example:controller}, and similarly for race-3-4x4.
For the bigger variants of race-3, \toolname fails to build the model for $mem=1$ due to its size.
On the other hand, the publicly available version of Inf-JESP cannot handle Dec-(PO)MDPs with more than two agents, thus no values are reported in the table.
For the variants available to both tools, one can see that the performance of \mbox{Inf-JESP} significantly lags behind \toolname both in the quality of the produced policies as well as the overall runtime.

\paragraph{Current state opacity}
Opacity with respect to the current state is a well-studied security/privacy planning requirement both in the deterministic~\cite{ZhangYZ19,YinL15, 0044NP20} and in the probabilistic setting~\cite{DobeSBBLPW23,AndriushchenkoBCPS23}.
It ensures that an intruder with limited observation capability is not able to fully guess an agent's current state.
We consider the variant of~\cite{DobeSBBLPW23}.
In our setting, two agents start from two close initial states and need to reach a target location via two different plans such that, in their synchronised runs, the agents always find themselves in the same region of the grid.
This way, the intruder cannot distinguish between the two. In \acs{PHLTL}:
\begin{gather*}
    \exists \, (\pv_1 \, \pv_2) \, . \, \forall \sv_{1} \in \{ s_1 \} (\pv_1) \, \forall \sv_{2}\in \{ s_2 \}(\pv_2) \, . \, \\
    \mathbb{P}_{\max} \big(\neg \llglob (\mathit{act}_{\sv_1} = \mathit{act}_{\sv_2}) \land \llglob (\mathit{reg}_{\sv_1} = \mathit{reg}_{\sv_2}) \land \lleven (\mathit{T}_{\sv_1}) \land \lleven (\mathit{T}_{\sv_2})\big)
\end{gather*}
where $\mathit{act}$ represents the action chosen by an agent in a state, and $\mathit{reg}$ is the current region of the agent (intruder's observation).
We observe that, for smaller variants, \toolname can quickly synthesise the best memoryless policy vector and can improve upon the obtained value when considering non-memoryless policies.
This is the only model in our benchmark set that seems to be favoured by Inf-JESP that can compute better solutions, although it fails to build a meaningful tuple of policies for the larger variant.
However, a high upper bound indicates that the policies computed by both tools may be far from the optimal one.

Up until now, the considered specifications belonged to the \acs{PHLTLDEC} fragment. We now consider problems that cannot be translated into a Dec-MDP formulation.

\paragraph{Initial state opacity (ISO)}
A variation of the opacity requirement is initial state opacity (ISO)~\cite{0044NP20,SabooriH13}.
Consider a variant where each action has some (different) probability of failing and going to a sink state $\mathit{S}$.
We seek one policy that is opaque with respect to the initial state, i.e. if the two agents execute the policy from two different initial states, they either simultaneously reach the target location or simultaneously end up in the sink state.
% Here the challenge is that in the synchronization of the two executions the agent may be in two different states, hence  taking two different actions, hence with different probability of failing.
In \acs{PHLTL}:
\vspace{-0.25em}
\begin{gather*}
    \exists \, (\pv_1) \, . \, \forall \sv_{1} \in \{ s_1 \} (\pv_1) \, \forall \sv_{2}\in \{ s_2 \}(\pv_1) \, . \, \\
    \mathbb{P}_{\max} \big( \, \lluntil{(\neg \mathit{T}_{\sv_1} \land \neg \mathit{T}_{\sv_2})}{(\mathit{T}_{\sv_1} \land \mathit{T}_{\sv_2})} \, \, \lor \\
    \lluntil{(\neg \mathit{S}_{\sv_1} \land \neg \mathit{S}_{\sv_2} \land \neg \mathit{T}_{\sv_1} \land \neg \mathit{T}_{\sv_2}\textbf{})}{(\mathit{S}_{\sv_1} \land \mathit{S}_{\sv_2})}\big)
\end{gather*}

\toolname can find the optimal memoryless policy for all variants but one (ISO-15x9) and, for the smallest variant (ISO-4x4), can even find the optimal policy with 1-bit memory.

\paragraph{Action failure robustness}
Robustness of the synthesized plan is a major concern in uncertain 
environments.
Here we consider action-failure robustness~\cite{0044NP20}.
We seek one robust policy for the initial state that, 
even in the case of a slip ($sl$)---%
the agent fails to execute an action and goes instead to a neighbouring cell---%
still manages to reach the target location as if 
the accident had not happened. Here the executions start from the same initial state. 
In \acs{PHLTL}:
\begin{gather*}
    \exists \, (\pv_1) \, . \, \forall \sv_{1} \in \{ s_1 \} (\pv_1) \, \forall \sv_{2}\in \{ s_1 \}(\pv_1) \, . \, \mathbb{P}_{\max} \Big( \big(\lleven \mathit{T}_{\sv_1} \land \lleven \mathit{T}_{\sv_2} \big) \land \\
    \big( (\lleven \, \mathit{sl}_{\sv_1} \oplus \lleven \, \mathit{sl}_{\sv_2}) 
    \implies \lluntil{(\neg \mathit{T}_{\sv_1} \land \neg \mathit{T}_{\sv_2})}{(\mathit{T}_{\sv_1} \land \mathit{T}_{\sv_2})} \big) \Big) 
\end{gather*}
where $\oplus$ is the \emph{xor} operator.
Regardless of the size, \toolname quickly synthesises the optimal memoryless policy with a value close to the upper bound. Adding more memory seems to have no effect.

\paragraph{Plan noninterference.}
With \acs{PHLTL} it is possible to express verification questions in path planning, such as \emph{noninterference}~\cite{DobeSBBLPW23,DimitrovaFT20}.
Again, two agents are moving in the grid towards a goal location from two initial locations far from each other.
The second one, A2, is an intruder that wants to interfere with A1 by reaching the goal first.
We want to verify whether changes in the policy of the intruder can affect A1. We seek one policy for A1 and two for A2 such that when the policy for A1 is executed together with one of the policies for A2, only in one execution the goal is accomplished. In \acs{PHLTL}:
%\vspace{-.5em}
\begin{gather*}
    \exists \, (\pv_{\mathit{A1}} \, \, \pv_{\mathit{A2a}} \, \, \pv_{\mathit{A2b}}) \, \,  . 
    \, \forall \sv_{\mathit{A1a}}\in \{ s_1 \} (\pv_{\mathit{A1}}) \,   \\ \forall \sv_{\mathit{A1b}}\in \{ s_1 \}(\pv_{\mathit{A1}}) \, \forall \sv_{\mathit{A2a}}\in \{ s_2 \}(\pv_{\mathit{A2a}}) \, \forall \sv_{\mathit{A2b}}\in \{ s_2 \}(\pv_{\mathit{A2b}}) \, . \\
    \mathbb{P}_{\max} \big( (\lluntil{\neg \mathit{goal}_{\sv_{\mathit{A1a}}}}{\mathit{goal}_{\sv_{\mathit{A2a}}}}) \oplus (\lluntil{\neg \mathit{goal}_{\sv_{\mathit{A1b}}}}{\mathit{goal}_{\sv_{\mathit{A2b}}}})\big) 
\end{gather*}
 We consider only the 4x4 variant of this problem as it requires a $4$-self-composition. \toolname quickly computes memoryless policies that show that the intruder can indeed interfere with A1 with a very high probability.

\section{Conclusion}
We introduce PHyperLTL, a novel specification formalism for decentralized planning under uncertainty.
It allows for specifying temporal constraints between policies and different executions of these policies by the agents. 
We use a synchronised product construction that allows us to (i) design an abstraction-refinement loop seeking the optimal joint policy and (ii) establish a close connection between a fragment of the logic and planning for Dec-MDPs. On a broad set of benchmarks, we demonstrate that the proposed abstraction-refinement approach can solve challenging problems in non-trivial environments and, in almost all cases, outperforms the state-of-the-art tool  \mbox{Inf-JESP} on Dec-MDP problems.

A major limitation of the proposed approach is the exponential blow-up of the state space yielding specifications that are sometimes intractable even for small environments.
In future work, we will investigate whether bottom-up approaches mitigating the state explosion~\cite{neary2021reward,schuppe2021decentralized} can be applied in our setting.

\clearpage

%%%%%%%%%%%%%%%%%%%%%%%%%%%%%%%%%%%%%%%%%%%%%%%%%%%%%%%%%%%%%%%%%%%%%%%%
%%%%%%%%%%%%%%%%%%%%%%%%%%%%%%%%%%%%%%%%%%%%%%%%%%%%%%%%%%%%%%%%%%%%%%%%

%%% The acknowledgments section is defined using the "acks" environment
%%% (rather than an unnumbered section). The use of this environment 
%%% ensures the proper identification of the section in the article 
%%% metadata as well as the consistent spelling of the heading.

\begin{acks}

% \begin{wrapfigure}{l}{0.5cm}
% \includegraphics[width=1cm]{eu_logo}
% \end{wrapfigure}
We are thankful to Tim Quatmann for some pointers to the Storm codebase.
\inlinegraphics{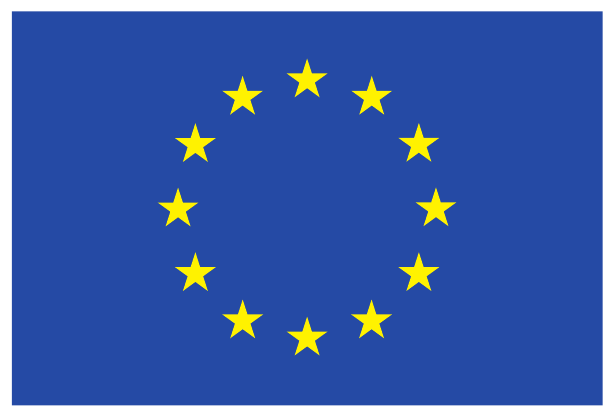} This work has been executed under the project VASSAL: ``Verification and Analysis for Safety and Security of Applications in Life'' funded by the European Union under Horizon Europe WIDERA Coordination and Support Action/Grant Agreement No. 101160022.
\inlinegraphics{eu_logo} This work has been partially funded by the  \grantsponsor{EUC}{EU Commission}{} in the Horizon Europe research and innovation programme under grant agreement No. \grantnum{EUC}{101034440} (Marie Sklodowska-Curie Doctoral Network LogiCS@TU Wien) and \grantnum{EUC}{101107303} (MSCA Postdoctoral Fellowship CORPORA).  Additionally, this work has been  funded by the Czech Science Foundation grant \mbox{GA23-06963S} (VESCAA), the IGA VUT project FIT-S-23-8151, and the \grantsponsor{WWTF}{Vienna Science and Technology Fund}{}  project \grantnum{WWTF}{ICT22-023}  (TAIGER).
\end{acks}

%%%%%%%%%%%%%%%%%%%%%%%%%%%%%%%%%%%%%%%%%%%%%%%%%%%%%%%%%%%%%%%%%%%%%%%%

%%% The next two lines define, first, the bibliography style to be 
%%% applied, and, second, the bibliography file to be used.

\bibliographystyle{ACM-Reference-Format} 
\bibliography{biblio}

\clearpage
\appendix

\section{Undecidability of \acs{PHLTL}}
\label{appendix:undec}

We report here the full proof of Theorem~\ref{thm:phltl-undecidable}.

Let $\mathcal{F} = (Q, \Sigma, \delta, q^0, q^\mathit{acc})$ be a \ac{PFA}~\cite{Paz71}, 
where $Q$ is a set of states, $\Sigma$ is the input alphabet, 
$\delta : Q \times \Sigma \rightarrow \distr(Q)$ is the transition probability function, 
$q^0 \in Q$ is the initial state, and $q^\mathit{acc} \in Q$ is an absorbing accepting state ($q^0 \neq q^\mathit{acc}$).
\ac{PFA} runs are defined analogously to \ac{MDP} paths by using input symbols as actions,
and the probability associated to a run is the product of the probabilities of all transitions it takes.
Let $\mathbb{P}_{\mathcal{F} (\omega)}$ be the probability 
that $\mathcal{F}$ is in $q^\mathit{acc}$ after reading string $\omega \in \Sigma^*$. 
The language accepted by a \ac{PFA} is defined with respect to a rational threshold $\lambda$: 
$L_{\lambda}(\mathcal{F}) = \{ \omega \in \Sigma^{+} \mid 
\mathbb{P}_{\mathcal{F}}(\omega) > \lambda \}$.
The emptiness problem for \acp{PFA} is known to be undecidable ~\cite{NasuH69,condon1989complexity,Freivalds81}.

\newtheorem*{theorem-undec}{Theorem 3.1}
\begin{theorem-undec}
    \acs{PHLTL} model checking is undecidable.
\end{theorem-undec}
\begin{proof}
Given $\mathcal{F}$ and $\lambda \in (0,1)$, we show how to construct 
a \ac{MDP} $M_\mathcal{F}$ and a formula $\Phi_\mathcal{F}$ such that 
$L_{\lambda}(\mathcal{F}) \neq \varnothing$ iff $M_\mathcal{F} \models \Phi_\mathcal{F}$.
We set $\Sigma_{\bot} = \Sigma \cup \{ \bot \}$.

$M_\mathcal{F}$ is the disjoint union of two \acp{MDP} $M_1$ and $M_2$.
Actions are \ac{PFA} input symbols, and action sequences of the two \acp{MDP} are strings read by $\mathcal{F}$.
$M_1$ mimics a run of $\mathcal{F}$, while keeping track of the last read symbol.
$M_2$ is a deterministic \ac{FSA} that just guesses $M_1$'s action sequence.
Formally, $M_\mathcal{F} = (S, \Act, \mpm, L)$ is defined as follows: 
\begin{itemize}
    \item % The states are
      $S = S_1 \cup S_2$,
      where $S_1 = \{ q_{\alpha} \mid q \in Q, \alpha \in \Sigma_{\bot} \}$
      contains $\mathcal{F}$'s states labeled with input symbols meant to track the last read symbol,
      while $M_2$'s state set $S_2 = \Sigma_{\bot}$
      consists of input symbols.
    \item $\Act = \{ \Sigma \mid \alpha \in \Sigma \}$:
      actions are shared among $M_1$ and $M_2$ and are meant to be synchronized. 
    \item The transition probability function $\mpm = \mpm_1 \cup \mpm_2$ is as follows:
    \begin{itemize}
        \item $\mpm_1(q_\alpha, \beta, p_{\beta}) = \delta(q, \beta, p)$
            for all $q,p \in Q$, $\alpha \in \Sigma_{\bot}$, $\beta \in \Act$,
            while $\mpm_1(q_\alpha, \beta, p_{\gamma}) = 0$ for each $\gamma \neq \beta$;
        \item $\mpm_2(\alpha, \beta, \beta) = 1$
            for all $\alpha \in \Sigma_{\bot}$ and $\beta \in \Act$,
            while $\mpm_2(\alpha, \beta, \gamma) = 0$
            for each $\gamma \neq \beta$;
    \end{itemize}  
    \item The atomic propositions are $AP = \Sigma_{\bot} \cup \{ t_{\alpha} \mid \alpha \in \Sigma \}$,
        where $t$-propositions mark the acceptance state $q^\mathit{acc}$.
        %are used to synchronize $M_1$ and $M_2$ at the end of the simulated \ac{PFA} run.
    \item The labeling function $L = L_1 \cup L_2$ is such that, for all $\alpha \in \Sigma_{\bot}$:
        \begin{align*}
            &L_1(q_{\alpha}) = \{ \alpha \} \text{ for $q \neq q^\mathit{acc}$} &
            &L_1(q^\mathit{acc}_{\alpha}) = \{t_\alpha \} \\
            &L_2(\alpha) = \{ \alpha \} &
            &
        \end{align*}
\end{itemize}

Formula $\Phi_\mathcal{F}$ is the following:
\begin{multline*}
\label{undec-formula}
\exists \, (\pv_1 \pv_2) \, . \, \forall \sv_{1} \in \{q^0_\bot \} (\sigma_1) \forall \sv_{2}\in \{ \bot \}(\sigma_2)  \, . \, \,  \\
\lprobsym \biggl( \Bigl( \bigvee_{\alpha \in \Sigma_{\bot}} \alpha_{\sv_{1}} \land \alpha_{\sv_{2}} \Bigr) \mathbin{\mathcal{U}} \Bigl( \bigvee_{\alpha \in \Sigma} {t_\alpha}_{\sv_{1}} \land {\alpha}_{\sv_{2}} \Bigr) \biggr) > \lambda
\end{multline*}

The formula considers two policies:
$\sigma_1$ controls $M_1$ and starts from $q^0_\bot$,
while $\sigma_2$ controls $M_2$ and starts from $\bot$.
The left-hand-side of the until operator forces them to take the same action at each time step,
while the right-hand-side prescribes 
$M_1$ to reach $q^\mathit{acc}$ (labeled by $t$), and that the last action played just before reaching $q^\mathit{acc}$ must be synchronized as well.
Thus, the action sequence taken by the two \acp{MDP} until $M_1$ reaches $q^\mathit{acc}$
can be seen as the string read by the \ac{PFA}.

We argue that it is possible to construct an accepting string for $\mathcal{F}$
from a pair of controllers $(\sigma_1, \sigma_2)$ satisfying formula $\Phi_\mathcal{F}$ on $M$, and \emph{vice versa}.
A planning algorithm computing them, or stating that they do not exist, 
would then solve the \ac{PFA} emptiness problem, which is undecidable.

Since $M_2$ is a deterministic \ac{FSA}, 
we can assume w.l.o.g.\ that $\sigma_2$ is deterministic, 
and corresponds to a finite string $\omega \in \Sigma^{+}$.
Since it satisfies the probability constraint of formula $\Phi_\mathcal{F}$,
there is a set of paths in $M_1$ that 
i) play actions $\nu \alpha$ for some prefix $\nu \alpha$ of $\omega$, and, following action sequence $\nu \alpha$, ii) reach $q^\mathit{acc}_{\alpha}$  with probability greater than $\lambda$. 
So, $\mathcal{F}$ accepts $\nu \alpha$ with probability greater than $\lambda$.
On the other hand, a string $\nu \alpha \in \Sigma^{+}$ 
that is accepted by $\mathcal{F}$ maps to two policies that are a witness for $\Phi_\mathcal{F}$.
Let $n = |\nu \alpha|$ be the length of such string.
Consider $\sigma_1 = \nu \alpha$ and $\sigma_2 = \nu\alpha$. 
These two policies are such that 
i) all paths in $M_1$ play the same actions as the only path in $M_2$; 
ii) the set of paths in $M_1$ that reach $q^\mathit{acc}_{\alpha}$ after $n$ steps has probability mass greater than $\lambda$.
iii) $M_2$ reaches $\alpha$ after $n$ steps. 
\end{proof}
The same result holds in case of infinite action sequences , or algorithms that create such finite or infinite strings such as finite-state controllers, and if we replace the strict inequality $>$ with $\geq$.

%%%%%%%%%%%%%%%%%%%%%%%%%%%%%%%%%%%%%%%%%%%%%%%%%%%%%%%%%%%%%%%%%%%%%%%%

\end{document}